# MEMORY EFFICIENT FORWARDING INFORMATION BASE FOR CONTENT-CENTRIC NETWORKING

Torsten Teubler[1], Dennis Pfisterer[2] and Horst Hellbrück[1]

[1]Lübeck University of Applied Sciences, CoSA Center of Excellence, Lübeck, Germany
torsten.teubler@fh-luebeck.de, horst.hellbrueck@fh-luebeck.de
[2]University of Lübeck, Lübeck, Germany
pfisterer@itm.uni-luebeck.de

## ABSTRACT

*Content-Centric Networking (CCN) is a new paradigm for the future Internet where content is addressed by hierarchically organized names with the goal to replace TCP/IP networks. Unlike IP addresses, names have arbitrary length and are larger than the four bytes of IPv4 addresses. One important data structure in CCN is the Forwarding Information Base (FIB) where prefixes of names are stored together with the forwarding face. Long prefixes create problems for memory constrained Internet of Things (IoT) devices. In this work, we derive requirements for a FIB in the IoT and survey possible solutions. We investigate, design and compare memory-efficient solutions for the FIB based on hashes and Bloom-Filters. For large number of prefixes and an equal distribution of prefixes to faces we recommend a FIB implementation based on Bloom-Filters. In all other cases, we recommend an implementation of the FIB with hashes.*

## 1. INTRODUCTION

We observe a trend on the Internet where content and services are more important than individual servers or hosts. Consequently, a new paradigm is required to address this development. Content-Centric Networking (CCN, cf. [1]) is one of these new paradigms where content is addressed by using hierarchically organized names with the goal on the long run to replace TCP/IP-based addressing. CCN, also known as named-data networking (NDN, cf. [2]) or information-centric networking (ICN, [3]) uses two important types of messages: interest and content object. In general, an interest message requests a content object. A major issue in CCNs is the forwarding of messages to the services that are mentioned/addressed by a certain interest message. Such services are addressed using names that are typically much longer than traditional addresses used in the Internet (e.g., IPv4 or IPv6 addresses). Instead of routing tables, CCN nodes contain a data structure called Forwarding Information Base (FIB) that backs the message forwarding process. However, due to the length of names, the memory consumption of FIBs is much higher than that of routing tables. As a result, CCNs may not be as scalable as IP- based networks. While these memory requirements are already an issue on standard PC hardware, it severely limits the applicability of CCN networking on highly resource-constraint devices in the Internet of Things (IoT). We address this issue in this paper by presenting the following contributions:

- We survey the requirements of Forwarding Information Bases in name and content-centric networks for the IoT.

- We define metrics for the requirements of the FIB.

- We develop and evaluate memory efficient solutions for a FIB based on hashes and counting and non-counting Bloom-Filters.

- We compare the solutions with a standard reference implementation of a FIB.

The remainder of the paper is organized as follows: Section 2 introduces some fundamentals on content-centric networks and introduces memory efficient data structures specific to this paper. Section 3 introduces related work; Section 4 presents the problem description. Section 5 introduces our memory efficient FIB implementations. In Section 6, we evaluate our memory efficient FIB implementations. Section 7 summarizes the work and gives an outlook on future research.

## 2. FUNDAMENTALS

This section introduces some prerequisites that are required to understand our approach. First, we discuss details of CCNs and addressing in CCNs. Second, we introduce some memory-efficient data structures that support addressing and routing in CCNs.

### 2.1. Content-Centric Networking

Each CCN message contains a name that is comprised of components separated by slashes (/). Components are strings of ASCII characters and bytes. Bytes not representing valid ASCII characters for a name are escaped (percent encoded) like in URLs. An example for a CCN name consisting of four components is /de/fhluebeck/ac/%C1.getTemp. %C1 is a percent encoded byte and some CCN implementations use this byte followed by a dot (%C1.) to annotate components with a special meaning. For example, %C1.getTemp is a remote method call to get a temperature.

Components starting from the beginning of a name are called prefixes of that name (just like network prefixes in IP addressing). Prefixes of the name above are, e.g., /de or /de/fhluebeck/ac.

The canonical ordering of names in CCNs is the shortlex order. A shortlex order is an alphabetical order on components, which additionally regards the length of the component. Shorter components are ordered before longer ones. The longest prefix match is important for name comparison. For example, the name /a/b/c has the prefixes /a and /a/b. Prefix /a/b matches /a/b/c better than /a according to the longest prefix match because two components from the beginning are equal.

Each CCN node has the following three data structures: forward-information base (FIB), content store (CS), and pending interest table (PIT). The FIB stores information about the interest forwarding, the CS caches received content objects, and the PIT stores received interests. Further building blocks of a node are faces and a daemon. A face is a generalization of an interface where interests and content objects are sent to and received from. The daemon processes the CCN messages.

If a node receives an interest, the contents of the CS get searched. The CS stores content objects by the canonical order of their names. If a matching content object is found the content object is returned to the face the interest was received from.

If the lookup in the CS was unsuccessful, the PIT is searched for a matching interest. The PIT stores an interest along with a list of faces the interest was received from. If the received interest matches an interest in the PIT the receiving faces for this entry are updated and the received interest is discarded.

If lookup in the PIT failed the FIB is searched. The FIB maps prefixes to faces. A received interest is compared to the prefixes in the FIB. The interest is forwarded to the faces with a longest prefix match. If an interest cannot be forwarded the interest is discarded finally.

If a node receives a content object with a matching interest in the PIT, the content object is forwarded to the faces where the interest was originally received from and stored in the CS.

The matching interest is removed from the PIT. Unlike IP addresses names are not limited in size and are typically larger in size than IPv4 addresses or IPv6 addresses.

Especially for the Internet of Things where severely resource constrained devices are integrated into the existing Internet an efficient representation of names and name prefixes is needed.

## 2.2. Memory-Efficient Data Structures

Unlike IP addresses, names have arbitrary length and are larger than e.g. four bytes of IPv4 addresses which increases the required memory. Well-known memory efficient representations for strings or names respectively, are hash values, Bloom Filters (BF) and Counting Bloom-Filters (CBF).

Hashing is a technique to map arbitrary length byte arrays (strings, objects, etc.) to a fixed size integer value by a hash function. The result of hashing is a hash value. Hash values are fixed in size and often smaller than the byte arrays that serve as input of the hash function. A so called hash collision occurs if hash functions return the same hash value on different inputs.

Wang and Kissel [4] explain the theory behind hashing and give recommendations to choose the hash value size to achieve a certain hash collision probability. Ahmad and Younis [5] present an evaluation of well-known hash functions. In our work, we assume uniformly distributed and efficiently computable hash functions. We apply hash functions in the context of CCN with the goal to minimize memory consumption.

A Bloom-Filter is a fixed size data structure to store a set of elements. By Bloom-Filters we can decide if an element is inside a set or not inside a set with a certain *probability*.

Especially in networking applications Bloom-Filters received broad attention. Broder and Mitzenmacher [6] survey network applications of Bloom-Filters and introduce the theory of BF and CBF in detail.

A Bloom-Filter is implemented as a bit array of size $m$. At insertion of an element, $k$ hash functions are applied to the element to be inserted. The hash functions compute index values in the range 1 to $m$. In the bit array, all bits with computed indices of the $k$ hash functions are set to one. The procedure described above is applied to all elements to be inserted into the Bloom-Filter. At each insertion it is possible that a bit in the array is chosen, which is already set to one. The false positive probability increases with the number of bits set in the Bloom-Filter. Therefore, the size of the Bloom-Filter (and the number of hash functions) is chosen so that it is able to hold a certain amount of elements and simultaneously maintaining a certain false positive probability. Note that, as with hashing, the size of the elements inserted into the BF does not matter.

It is not possible to remove an element from a Bloom-Filter. Trying to do so would probably remove other elements as well. So called counting Bloom-Filter (CBF) [7] allow removing elements. Counting Bloom-Filters are an extension of BF where the bits of the Bloom-Filter are replaced by counters. At each insertion, the counters are incremented accordingly in a counting

Bloom-Filter and decremented if an element is removed. Fan et al. [7] show that counters with four bits are sufficient for most applications.

## 3. RELATED WORK

In this section, we introduce relevant work and state of the art in hashing and Bloom-Filters and their usage in content-centric networking. In CCN there are plenty of applications for hashing and Bloom-Filter due to the names with variable lengths.

Hashing is widely used in networking applications. Saino et al. [8] present an example for hashing in information-centric networking (ICN). They propose a solution to the routing problem in information-centric networks e.g. the shortest path to the router which temporarily hosts the requested information or content. Regarding the routing problem is not in the scope of our work but nevertheless, our solution with hashing is flexible and allows the manipulation of the FIB to enable a routing mechanism on top of it.

Varvello et al. [9] develop a content router for high speed forwarding. They use Bloom-Filter for longest prefix matches. Unlike our solution, their aim is to decrease the time needed for forwarding messages on routers while decreasing memory usage is not in their focus. Furthermore, their longest prefix match algorithm is embedded in a complex, heavyweight architecture.

Wei You et al. [10] propose Bloom-Filters to decrease memory usage of the PIT. In their approach, each face has a counting Bloom-Filter and therefore, they call it "distributed Bloom-Filter". The counters of the Bloom-Filters are incremented on interest reception and decremented on content object forwarding. Furthermore, they introduce a shared Bloom-Filter which is applied after the distributed Bloom-Filters to further reduce the false positive results. In contrast, we propose a simpler design and do not create a sequence of Bloom-Filters. In addition, we suggest using a single Bloom-Filter per face.

Tsilopoulos et al. [11] also propose Bloom-Filters to reduce the space needed for content forwarding information in the PIT. In their approach the interests are tracked only at a subset of hops. Intermediate nodes not tracking the interests use a Bloom-Filter at each face to store the forward information. This application tolerates high false-positive rates of the Bloom-Filters because the falsely forwarded interests are stopped at nodes tracking the whole interest in their PIT. Therefore, the Bloom-Filters require less memory in that application.

Wei Quan et al. [12] propose a hybrid approach using Tree-Bitmaps and Bloom-Filters for a scalable name lookup in CCNs. They split names are split in two segments. The first segment is a fixed size part of a name and is processed by the Tree-Bitmap. The second part is of variable size and is processed by the Bloom-Filter. Wei Quan et al. compare name lookup speed and memory usage of their approach to an alternative "Name Prefix-Trie" implementation.

An interesting approach for a longest prefix match on IP addresses with Bloom-Filters is described by Dharmapurikar et al. [13]. Their strategy for longest prefix matching is to use one Bloom-Filter per prefix. We use one Bloom-Filter per face and store the prefixes as elements in the Bloom-Filter.

The work of Muñoz et al. [14] is closest to our Bloom-Filter approach regarding aim and scope. They suggest so called *iterated* Bloom-Filters for use in resource constrained IoT devices in content-centric networks. Like in our solution they use one (iterated) Bloom-Filter per face.

The advantage of iterated Bloom-Filters it that they either allow the reduction of the Bloom-Filter size while maintaining a certain false positive probability or a reduction of the false positive probability by maintaining the size of the Bloom-Filter.

Muñoz et al. discuss the properties of iterated Bloom-Filters comprehensively but the work lacks a detailed discussion of the removal of elements from the Bloom-Filters by using counting Bloom-Filters. Furthermore, it is questionable if their approach really fits the requirements of resource constrained devices because the iterated Bloom-Filters still become quite large (several kilo bytes up to several giga bytes in their evaluation).

Compared with the solution of Muñoz et al. our approach with Bloom-Filters is simpler and probably less efficient but allows for better maintenance and error handling of Bloom-Filters in content-centric networking. Furthermore, we show that the mapping of prefixes to faces has an influence when each face has its own Bloom-Filter. We discuss and investigate counting Bloom-Filters which are important for dynamic FIBs and compare the approach with a lightweight solution using hashing. Our implementation targets resource-constraint IoT devices.

## 4. PROBLEM DESCRIPTION

We survey the requirements for FIBs and introduce estimations for the FIB size. The estimated FIB size is one basic parameter for our evaluation presented in Section 6.

From a data structure perspective, an FIB maps a prefix to multiple faces. The relation between prefixes and faces presented in Figure 1 shows that a prefix has at least one face it is assigned to. In general, a face is not required to have a prefix assigned to it.

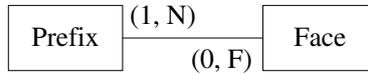

Figure 1. ER-diagram showing the relation between prefix and face.

We have implemented a FIB for microcontroller-based IoT devices, which is already memory efficient by using a bit vector for the list of faces. This implementation serves as a reference for our alternative approaches. Figure 2 shows the structure of our FIB implementation.

The FIB is a sorted list of $N$ tuples of a prefix and a bit vector ($[b_1, b_2, ..., b_F]$). The list is sorted according to the canonical order of the prefixes for a more efficient lookup. Prefix $n$ consists of $M_n$ components ($Comp_{n,M_n}$) where $n \in \{1, ..., N\}$. Each component is a byte string and the components are separated by a one byte sized type/length field (cf. our previous work in [15]). If a prefix is assigned to the $f$-th face, the $f$-th bit ($b_f$) in the corresponding bit vector is set to one, where $f \in \{1, ..., F\}$ and $F$ is the number of faces of that node.

In the following, we compare the size of this reference implementation of the FIB with hashing and Bloom-Filter solutions. The size of Bloom-Filters is usually given in bits. For a quantitative comparison, the size of our FIB implementation in bit is calculated as

$$\sum_{n=1}^{N} \sum_{m=1}^{M_n} (8 \cdot \left| Comp_{n,m} \right|_{Byte} + 1) + F) \qquad (1)$$

where $\left| Comp_{n,m} \right|_{Byte}$ is the size in byte of the $m$-th component of the $n$-th prefix.

Regarding individual components of a prefix is not necessary for our comparison in Section 6 because we regard the prefix as a whole. We denote the whole prefix as $Prefix_n$ (including separators) for a FIB entry. Table 1 summarizes the symbol definitions used throughout this work.

Table 1. Symbol definitions.

| Symbol | Description |
|---|---|
| $F$ | Number of faces per node |
| $N$ | Number of prefixes in the FIB |
| $M_n$ | Number of components of the $n$-th FIB prefix |
| $Comp_{n,m}$ | $m$-th component of the $n$-th FIB prefix |
| $Prefix_n$ | The $n$-th FIB prefix ($Comp_{n,1}$ ... $/Comp_{n,M_n}$) |
| $\|\cdot\|_{Bit}$ or $\|\cdot\|_{Byte}$ | Size in bit or byte of $\cdot$ |

In practice, relations between prefixes and faces may show different characteristics depending on the number of prefixes $N$ and the number of faces $F$. These characteristics define the requirements for FIBs and are important for choosing the best optimization strategy. Our reference implementation for the FIB (cf. Figure 2) for example grows faster in size with increasing $N$ than with increasing $F$. The length of each prefix |Prefix n | also has a significant influence on the size of our reference implementation (cf. Figure 2) but not for our suggested optimization approaches. Therefore, we further simplify our considerations by assuming that every prefix in the FIB has a fixed size, denoted by $|Prefix_n|_{Bit}$ or $|Prefix_n|_{Byte}$, respectively. According to this assumption, equation (1) simplifies to equation (2).

$$N \cdot (|Prefix_N|_{Bit} + F) \tag{2}$$

Regarding the parameters N and F, we identify three cases: 1. $N \approx F$, 2. $N \ll F$, and 3. $N \gg F$. According to these three cases, relations as illustrated in Figure 1 lead to three solutions for the mapping as given in Figure 3.

For a 1-1-mapping one prefix maps to one face, 1-n one prefix maps to many faces, and n-1 many prefixes map to one face. Our mapping characteristics shown in Figure 3 represent "extreme" cases. In practice the "1" in our mapping characteristics is not required to be exactly one, mostly one suffices. For example, the 1-1-mapping maps mostly one prefix to mainly one face, and so on.

It is hard to answer which mapping characteristic usually applies to FIBs in CCNs as real-world implementations are rare. CCN and applications on top of it are still developing and surveys are—to the best of our knowledge—not available.

$$
\begin{aligned}
&(/Comp_{1,1} \quad /Comp_{1,2} \quad \cdots \quad /Comp_{1,M_1}, \quad [b_1, b_2, \ldots, b_F]), \\
&(/Comp_{2,1} \quad /Comp_{2,2} \quad \cdots \quad /Comp_{2,M_2}, \quad [b_1, b_2, \ldots, b_F]), \\
&\vdots \quad\quad \cdots \quad\quad \ddots \quad\quad \vdots \quad\quad \vdots \\
&(/Comp_{N,1} \quad \cdots \quad /Comp_{N,M_N-1} \quad /Comp_{N,M_N}, \quad [b_1, b_2, \ldots, b_F])
\end{aligned}
$$

Figure 2. Structure of our reference FIB implementation or the IoT.

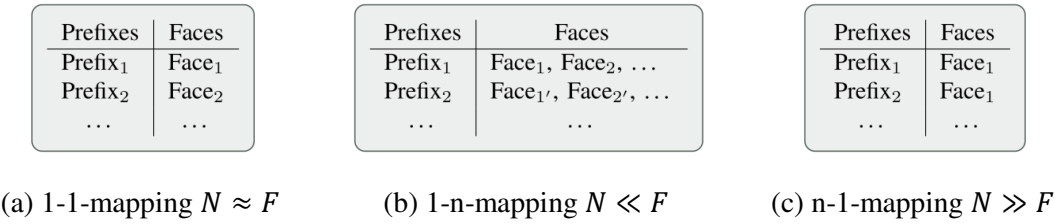

(a) 1-1-mapping $N \approx F$         (b) 1-n-mapping $N \ll F$         (c) n-1-mapping $N \gg F$

Figure 3. FIB mapping characteristics.

Therefore, we surveyed FIBs of existing testbeds to get an impression of the characteristics and to fixing values for our evaluation in Section 6. We conducted measurements using the NDN testbed (https://named-data.net/ndn-testbed/) with 28 nodes (cf. Figure 4) and we also conducted measurements of our content-centric networking implementation for the IoT.

Figure 4a presents survey results for the NDN testbed. Each column in Figure 4a represents an NDN node. Labels of the nodes (x axis) are shown on Figure 4b. The height of a column is the number of faces $F$ of the node. Different segments of the columns in Figure 4a show the number of faces with a certain number of prefixes assigned to it. E.g., where 0, between 1 and 4, between 5 and 7, between 10 and 11, and between 56 and 62 prefixes, are assigned to a face. For example, the first node (/ndn/br/ufpa) has 5 faces where no prefix is assigned and 2 faces where between 56 and 62 prefixes are assigned to. The upper x-axis shows the number of prefixes $N$ in the FIB of that node.

Figure 4b is similar to Figure 4a but the y-axis shows the number of prefixes $N$ and the upper x-axis the number of faces $F$. The different segments of the columns in Figure 4b show the number of prefixes assigned to 1 and 2, 3 and 4, 5 and 6, 7 and 8, and 9 and 10 faces, respectively. The first node in Figure 4b has more than 70 prefixes which are assigned to 1 or 2 faces only.

If $N \approx F$ and the values in the columns are small (near to one), then it is a 1-1-mapping. If $N \ll F$ and the values in the columns in the first graph (Figure 4a) are small and in the second graph (Figure 4b) are large, then it is a 1-n-mapping. If $N \gg F$ and the values in the columns of the first graph are large and in the second are small, then it is an n-1-mapping.

A classification in one of the mapping types shown in Figure 3 is difficult in Figure 4. One subset of prefix to face mappings in Figure 4 indicates a 1-1-mapping, whereas another subset indicates an n-1-mapping.

Figure 5 shows the FIB characteristics for our content-centric networking IoT implementation in a plot similar to Figure 4. We have several IoT nodes with an identical FIB configuration and therefore, one column (IoT-Nodes) represents all nodes. The second column represents the FIB characteristics for the IoT gateway (IoT-GW). Our content-centric IoT gateway is responsible for the wired/wireless transition and has more resources than IoT nodes.

Figure 5a and Figure 5b show that the FIB in our CCN testbed setup for the IoT follows a 1-1-mapping characteristic.

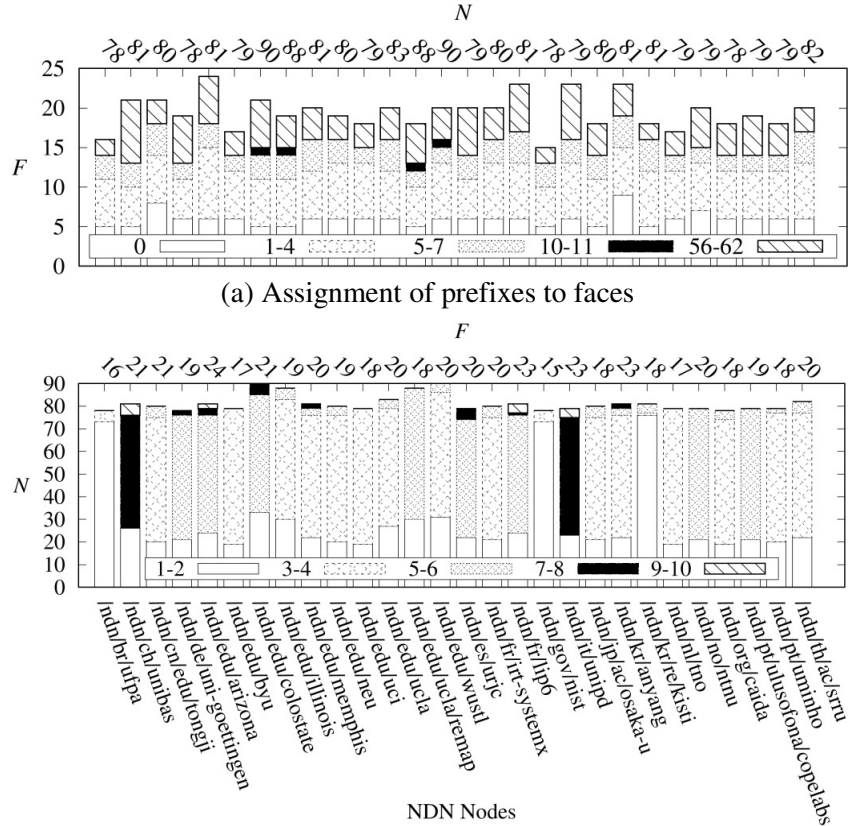

(a) Assignment of prefixes to faces

(b) Assignment of faces to prefixes

Figure 4. FIB characteristics of the NDN testbed.

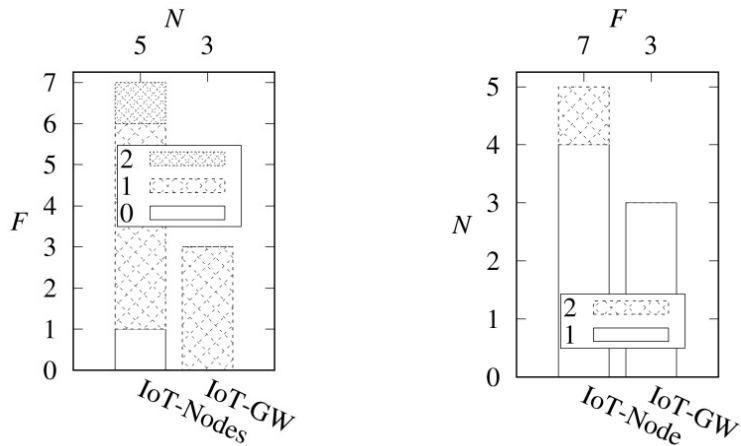

(a) Assignment of prefixes to faces    (b) Assignment of faces to prefixes

Figure 5. FIB characteristics of the IoT nodes and the IoT gateway in our testbed.

In summary, our survey gives an indication how FIB characteristics develop. In our opinion, FIB in future content-centric networking will show 1-1 or n-1-mappings. Mapping characteristics are important for choosing the correct optimization strategy. For example, in Section 5 and Section 6 we show, that Bloom-Filters work efficiently for the n-1-mapping characteristic.

## 5. MEMORY EFFICIENT FIB IMPLEMENTATIONS

In this section, we present alternative solutions for the FIB with the aim to reduce memory consumption. The solutions presented here rely on hashing and Bloom-Filters. In the following, our reference FIB implementation (cf. Figure 2) is named FIB, the FIB with hashing is named FIB-Hash, and the FIB solution using Bloom-Filters is called FIB-BF or FIB-CBF with counting Bloom-Filters, respectively.

### 5.1. FIB Implementation with Hashes

Compared to the reference implementation of the FIB, the FIB-Hash implementation prefixes are replaced by hash values as shown in Figure 6. As a consequence, the longest prefix matching algorithm in FIB-Hash is different from the reference FIB.

On reception of an interest, a hash value is computed from the name of the interest and compared to hash values 1 to $N$ in FIB-Hash. If the hash value of the interest is equal to a FIB-Hash hash value, the interest is forwarded to the faces defined in the bit vector and the FIB processing stops. If no matching hash value in FIB-Hash is found, the hash value of the name excluding the last name component is computed and the comparison starts from the beginning. With the first matching hash entry the processing stops. A successful comparison is always a longest prefix match because the remainder of the interest name is the longest prefix of the interest name in the current round. Figure 7 shows the interest name processing in FIB-Hash by an example.

In the example, the interest name has four components and the FIB-Hash manages three prefixes ($N = 3$). The first step (1.) depicts the name processing in FIB-Hash. In the first round, a hash value is computed from the whole name, in the second of the first three components, and in the third round of the first two components. The hash value 9c4e matches a forwarding entry in FIB-Hash and the interest is forwarded to the corresponding faces of that forwarding entry in the second step (2.).

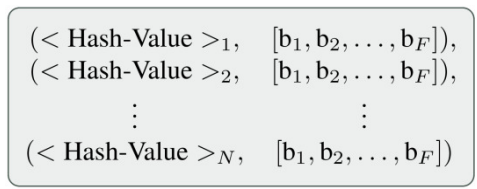

Figure 6. Structure of our FIB-Hash implementation.

We denote the probability that a prefix is erroneously identified as a member in FIB-Hash as false-positive probability ($P_{FP}$). According to a simplified approach by Mitzenmacher and Broder [6] the maximum false-positive probability is approximately $10^{-5}$ for $N = 2^{16}$ and a four byte hash value. The false-positive probability of $10^{-5}$ is a large-scale upper bound for the real false-positive probability, which is smaller by magnitudes because we assume $N$ between 50 and 100. According to equation (2) the size in bit for our FIB-Hash is $N(32 + F)$ when using four byte hash values.

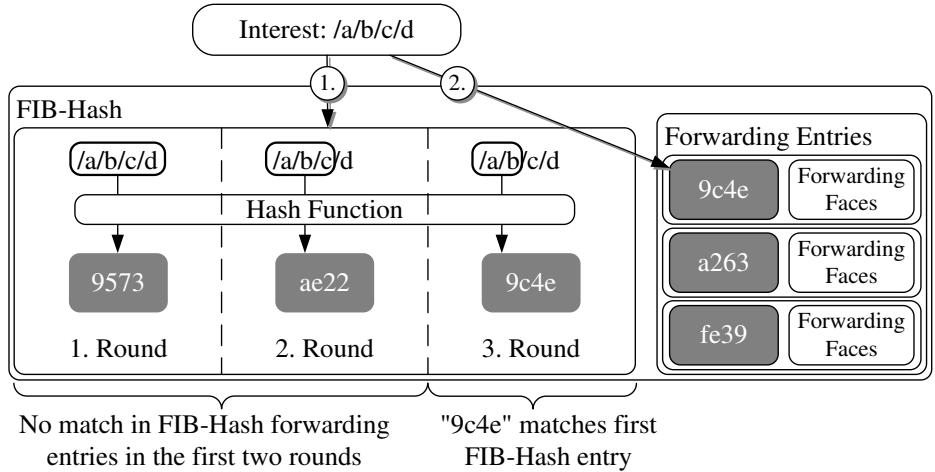

Figure 7. FIB-Hash interest processing.

For efficiency reasons, an implementation may compute the hash iteratively component by component starting from the first component. The intermediate results will be stored. The comparison against the FIB entries will also start with the hash of the whole name and continued with the intermediate results to apply the hashes like in Figure 7 until a match is found.

## 5.2. FIB Implementation with Bloom-Filters

Figure 8 shows the structure of FIB-BF. In contrast to the FIB-Hash implementation (where a hash value is assigned for each prefix), FIB-BF assigns a Bloom-Filter to each face. Each Bloom-Filter contains the prefixes assigned to that face. The interest processing is similar to the interest processing of FIB-Hash. Each prefix (from long to short) is checked if it is contained in one of the F Bloom-Filters. Unlike to FIB-Hash where the processing stops at first match the prefix is checked against all Bloom-Filters because other faces might have this prefix assigned, too.

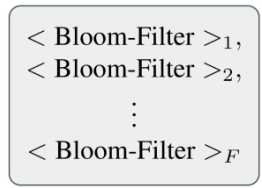

Figure 8. Structure of our FIB-BF implementation.

All Bloom-Filters shown in Figure 8 have the same size. In general, the size of a Bloom-Filter depends on the desired false-positive probability $P_{FP}$ and the maximum capacity of the Bloom-Filter $N_{BF}$ for which $P_{FP}$ holds. It is possible to add more elements than a Bloom-Filters capacity, but then the desired $P_{FP}$ does not hold anymore.

Let $size_{BF}(N_{BF}, P_{FP})$ the size of a Bloom-Filter in bit (depending on $N_{BF}$ and $P_{FP}$). The size in bit of FIB-BF is then given by equation (3).

$$size_{BF}(N_{BF}, P_{FP}) \cdot F \qquad (3)$$

We require $P_{FP} = 10^{-6}$ and a nearly equal distribution of prefixes among faces, hence, $N_{BF} \approx N/F$. If the latter requirement is not met FIB-BF may end up with Bloom-Filters containing fewer elements than its capacity which is inefficient memory usage.

The design of our FIB-Bloom in Figure 8 suggests an n-1-mapping because the Bloom-Filters in FIB-Bloom include multiple prefixes assigned to one face. If we regard a FIB-BF with fixed $N_{BF}$ or $F$ respectively, the memory consumption of FIB-BF increases faster with increasing $F$ than with increasing $N_{BF}$. Therefore, FIB-BF requires an n-1-mapping with a few faces in total.

### 5.3. FIB Implementation with Counting Bloom-Filters

The FIB implementation with counting Bloom-Filter FIB-CBF is similar to FIB-BF discussed in Section 5.2. Counting Bloom-Filters replace the single bits with fixed size counters. Therefore, the size of the FIB-CBF compared to FIB-BF increases by a fixed amount. Counting Bloom-Filters allow—in contrast to Bloom-Filters—the removal of elements. Removal of prefixes is necessary when forwarding information needs to be adapted at runtime.

For our FIB-CBF we use counters with a size of four bits according to recommendation of Fan et al. [7]. With FIB-Hash and FIB-BF/CBF we introduce new symbols summarized in Table 2 which completes Table 1.

Table 2 Symbol Definitions for FIB-HASH and FIB-BF/CBF.

| Symbol | Description |
|---|---|
| $P_{FP}$ | False-positive probability |
| $N_{BF}$ | Bloom-Filter capacity |
| $size_{BF}(N_{BF}, P_{FP}) \cdot F$ | Bloom-Filter size in bit |

## 6. EVALUATION

In this section, we compare the memory consumption of our FIB implementation and our alternative solutions FIB-Hash and FIB-BF/CBF. We determine the memory consumption by conducting a theoretical analysis under assumption of reasonable values for $N$, $F$, $N_{BF}$, $|\text{Prefix}_N|_{Byte}$, and $P_{FP}$. In the following, we first introduce our evaluation setup. Second, we evaluate the influence of $N$, and $F$. Third, we evaluate the influence of the prefix length $|\text{Prefix}_N|_{Byte}$.

### 6.1. Evaluation Setup and Assumptions

In the NDN testbed (cf. Figure 4) the number of prefixes $N$ is around 80 per FIB. For our future content-centric IoT setup we expect about 30 prefixes per FIB. Therefore, we assume $N$ around 50 to 60 in our evaluation setup because it is in the middle of the range from 30 to 80. Assuming a higher value for $N$ provides advantages to FIB-Hash and FIB-BF/CBF memory consumption because memory consumption of our standard FIB implementation performs worse with increasing number of prefixes $N$.

In the NDN testbed (cf. Figure 4) each node has 15 to 24 faces and at least 5 faces per node have no prefix assigned. For our future content-centric IoT setup we expect a node to have 5 to 20 faces. Therefore, it seems reasonable to assume a number of faces $F$ between 10 and 15 in our evaluation setup.

For the prefix length $|\text{Prefix}_N|_{Byte}$, we assume 15 byte. In the NDN testbed we observe two accumulations of prefix lengths shown in Figure 9.

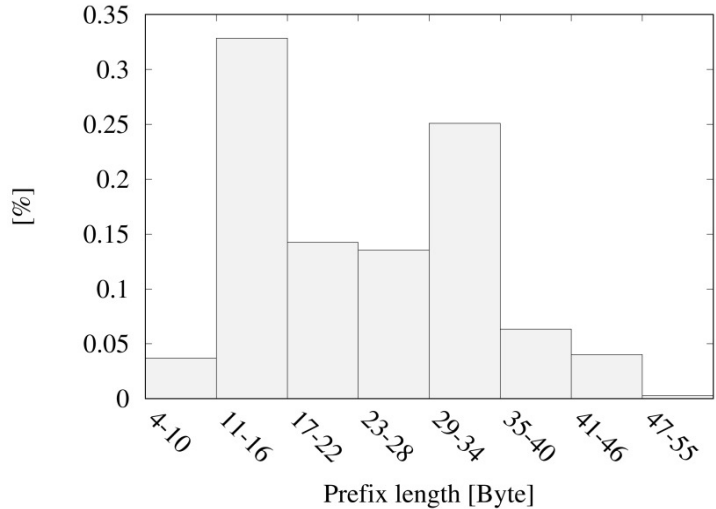

Figure 9 Histogram of the prefix lengths in the NDN testbed.

One cluster is at around 15 byte and the other cluster is at around 30 byte. A prefix length of 15 byte also is reasonable for our future content-centric IoT setup. Therefore, we choose $|\text{Prefix}_N|_{Byte} = 15$. Larger prefixes provide advantages for FIB-Hash and FIB-BF/CBF memory consumption because with FIB-Hash and FIB-BF/CBF the prefix length has no influence (cf. equation (3)) but for our standard FIB implementation (cf. equation (2)). In our evaluation setup we assume a false-positive probability for FIB-BF/CBF of $10^{-6}$. This false-positive probability provides a reasonable trade-off between the Bloom-Filter size and precision. It is quite obvious that FIB-Hash performs better than the standard FIB implementation regarding the memory consumption and therefore, we focus on FIB and FIB-BF/CBF in the following.

Table 3 summarizes our evaluation setup. We define $N_{BF}$ depending on $N$ and $F$ according to the observations in Section 5.2 and we apply the ceiling function to ensure that $N_{BF}$ is an integer allowing FIB-BF/CBF to store all $N$ prefixes. Unless stated otherwise, the values in Table 3 hold for the evaluation setup in this section.

Table 3 Evaluation setup.

| Symbol | Value(s) |
|---|---|
| $N$ | 50 or 60 |
| $F$ | 10 to 15 |
| $N_{BF}$ | $\lceil N/F \rceil$ |
| $|\text{Prefix}_N|_{Byte}$ | 15 |
| $P_{FP}$ | $10^{-6}$ |

## 6.2. Influence of *N* and *F*

Figure 10 shows how the size of FIB and FIB-BF/CBF changes with increasing number of faces. The x-axis shows the number of faces $F$ and the y-axis the size of the corresponding data structure in kbits.

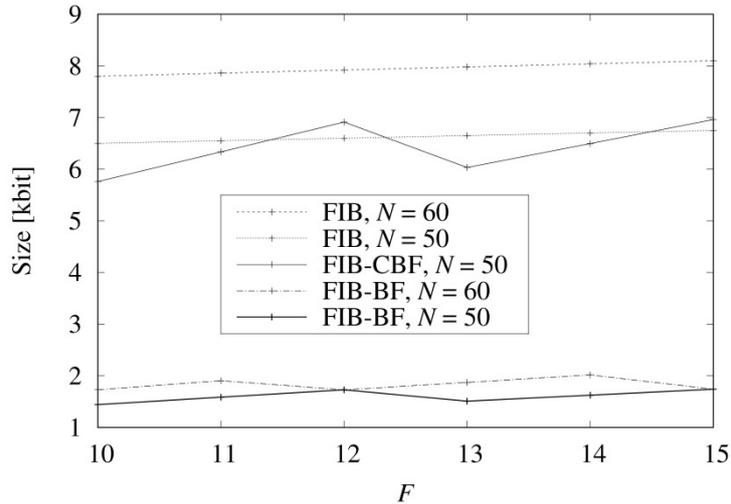

Figure 10. Size comparison of FIB and FIB-BF/CBF.

Table 4 shows the detailed results of Figure 10. Under the assumption that all prefixes are distributed equally among the faces FIB-BF consumes considerably less memory than FIB. In contrary to FIB, FIB-BF size is always in bounds with increasing $F$. However, if we increase $F$ so that $N = F$ we end up with a 1-1-mapping which suits better for FIB-Hash. We omit results for FIB-CBF $N = 60$ in Figure 10 as memory size is too large.

Table 4 Detailed results of Figure 10.

| FIB Type | N | Memory Consumption [kbit] | | | | | |
|---|---|---|---|---|---|---|---|
| | | $F = 10$ | $F = 11$ | $F = 12$ | $F = 13$ | $F = 14$ | $F = 15$ |
| FIB | 60 | 7.8 | 7.86 | 7.92 | 7.98 | 8.04 | 8.10 |
| | 50 | 6.5 | 6.55 | 6.6 | 6.65 | 6.7 | 6.75 |
| FIB-CBF | 50 | 5.76 | 6.336 | 6.912 | 6.032 | 6.496 | 6.96 |
| FIB-BF | 60 | 1.73 | 1.903 | 1.728 | 1.872 | 2.016 | 1.74 |
| | 50 | 1.44 | 1.584 | 1.728 | 1.508 | 1.624 | 1.74 |

Also, the distance between FIB-BF $N = 50$ and $N = 60$ is lower than for FIB $N = 50$ and $N = 60$. Table 5 shows the distance (difference between $N = 50$ and $N = 60$) the curves (e.g. in Table 4: $7.8 - 6.5 = 1.3$). For FIB the difference is about 1.3 kbit and for FIB-BF between 0 and 0.39 kbit. In summary, FIB-BF scales better with increasing $N$ compared to FIB.

Table 5 Distance (difference) between curves $N = 50$ and $N = 60$ for FIB and FIB-BF.

| FIB Type | Difference [kbit] | | | | | |
|---|---|---|---|---|---|---|
| | $F = 10$ | $F = 11$ | $F = 12$ | $F = 13$ | $F = 14$ | $F = 15$ |
| FIB | 1.3 | 1.31 | 1.32 | 1.33 | 1.34 | 1.35 |
| FIB-BF | 0.29 | 0.319 | 0 | 0.364 | 0.392 | 0 |

The assumption that prefixes are equally distributed among the faces and $N_{BF} \approx {N}/{F}$ is important for FIB-BF/CBF. Figure 11 shows the result if the requirement is not met. In the setup in Figure 11 $N$ is 50 and $N_{BF}$ is 5, 10, and 25, respectively. $N_{BF} = 5$ correlates to $N_{BF} \approx {N}/{F}$ in the range of 10 to 15 faces. With increasing $N_{BF}$ the memory consumption of FIB-BF increases.

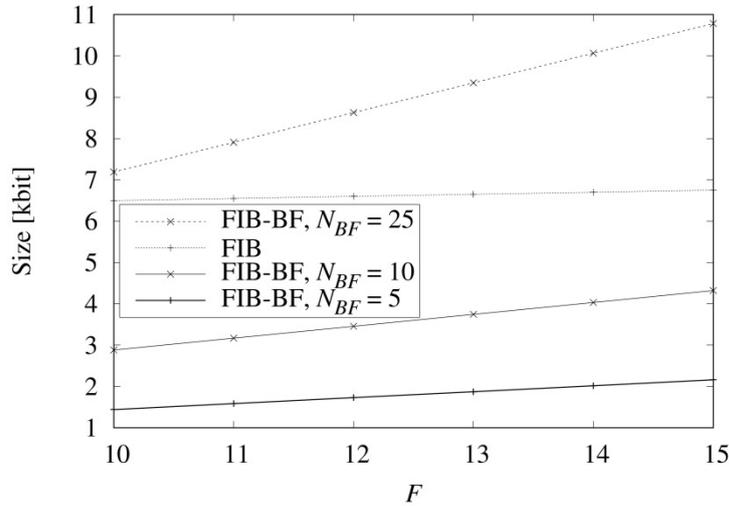

Figure 11. Size of FIB and FIB-BF/CBF and different values of the Bloom-Filter capacity $N_{BF}$.

Table 6 shows the detailed results of Figure 11. Note that the second column of Table 6 shows the capacity of the Bloom-Filter $N_{BF}$ and not the number of prefixes $N$. In the whole range of 10 to 15 faces there are no memory savings of FIB-BF with $N_{BF} = 25$ compared to FIB.

Table 6 Detailed results of Figure 11.

| FIB Type | $N_{BF}$ | Memory Consumption [kbit] | | | | | |
|---|---|---|---|---|---|---|---|
| | | $F = 10$ | $F = 11$ | $F = 12$ | $F = 13$ | $F = 14$ | $F = 15$ |
| FIB | N/A | 6.5 | 6.55 | 6.6 | 6.65 | 6.7 | 6.75 |
| | 25 | 7.19 | 7.909 | 8.628 | 9.347 | 10.066 | 10.785 |
| FIB-BF | 10 | 2.88 | 3.168 | 3.456 | 3.744 | 4,032 | 4,32 |
| | 5 | 1.44 | 1.584 | 1.728 | 1.872 | 2.016 | 2.16 |

## 6.3. Influence of prefix length

In the last part of the evaluation we investigate the influence of the prefix length on the memory consumption. With increasing prefix length FIB-BF and FIB-CBF outperform our reference FIB. Figure 12 shows the minimum prefix length (on the y-axis) where FIB-BF/CBF outperforms FIB regarding memory consumption for varying F (on the x-axis). $N$ is 50 in the setup shown in Figure 12.

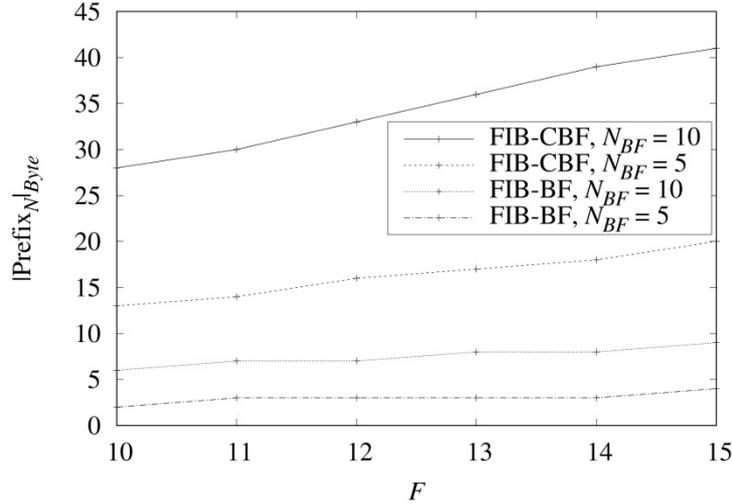

Figure 12. Minimum prefix length for different Bloom-Filter capacity $N_{BF}$ with number of prefixes $N = 50$ where FIB-BF/CBF outperforms FIB regarding memory consumption.

Table 7 shows detailed results of Figure 12. Even for short prefixes FIB-BF saves memory compared to FIB. If $N_{BF}$ increases from 5 to 10, the prefix length for which FIB-BF saves memory also increases by approximately the factor of two. For FIB-CBF the prefix has to be approximately 4 to 6 times longer so that FIB-CBF saves memory compared to FIB. In our opinion, prefix lengths for forwarding entries exceeding 30 byte seem unrealistic for us, even for future applications.

Table 7 Detailed results of Figure 12.

| FIB Type | $N_{BF}$ | $|\text{Prefix}_N|_{Byte}$ | | | | | |
| --- | --- | --- | --- | --- | --- | --- | --- |
| | | $F = 10$ | $F = 11$ | $F = 12$ | $F = 13$ | $F = 14$ | $F = 15$ |
| FIB-CBF | 10 | 28 | 30 | 33 | 36 | 39 | 41 |
| | 5 | 13 | 14 | 16 | 17 | 18 | 20 |
| FIB-BF | 10 | 6 | 7 | 7 | 8 | 8 | 9 |
| | 5 | 2 | 3 | 3 | 3 | 3 | 4 |

Figure 13 summarizes the behaviour of FIB, FIB-Hash, and FIB-BF/CBF. The y-axis in Figure 13 shows the quotient of the sizes of FIB-CBF, FIB-Hash, and FIB-BF to the FIB size on the y-axis against the prefix length on the x-axis. We call this quotient size ratio in the following. If the size ratio is below 1 (dashed horizontal line) FIB-Hash, FIB-BF, and FIB-CBF memory consumption is lower than for our reference FIB implementation. The cases for FIB-BF, $N = 30$, $F = 10$, $N_{BF} = 6 = {}^{2N}\!/_F$ and $N_{BF} = 9 = {}^{3N}\!/_F$ (Figure 13a), respectively, show the behaviour of FIB-BF if prefixes are not distributed equally among faces.

With increasing $F$ the FIB-BF size increases more than with increasing $N$. Table 8 shows this behavior on selected values from Figure 13. In Table 8 prefix lengths with different N are more similar than with different F.

FIB-Hash prefixes longer than 4 byte save memory because we suggest 4 byte hash values. FIB-CBF with $N = 30$ and $F = 10$ saves memory if prefixes are longer than 14 byte. FIB-CBF size is always larger than FIB-Hash.

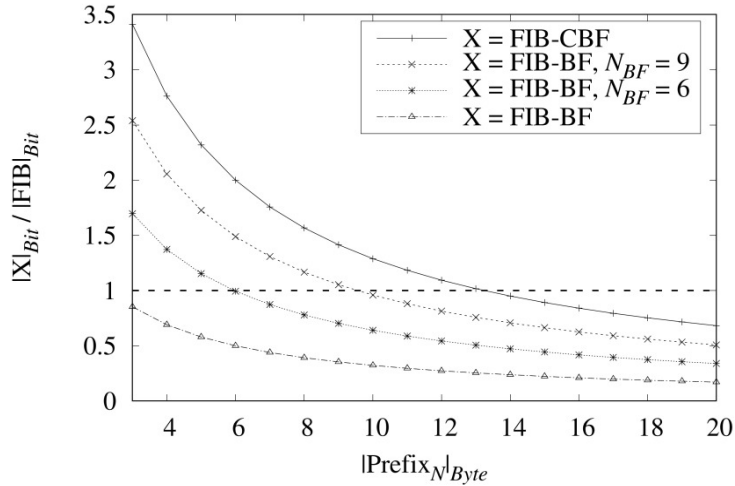

(a) $N = 30$, $F = 10$

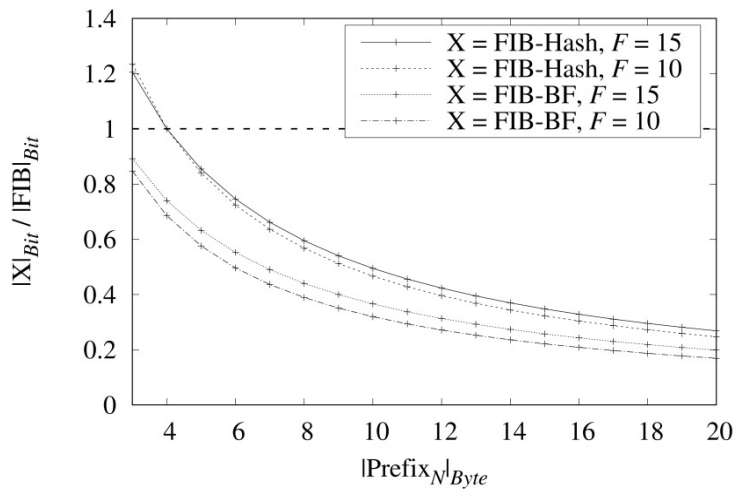

(b) $N = 50$

Figure 13. Prefix length against size ratio (FIB-Hash, FIB-BF/CBF size divided by FIB size).

Table 8 Selected values from Figure 13.

| Label | | $|\text{Prefix}_N|_{Byte}$ | | | |
|---|---|---|---|---|---|
| | | 4 | 6 | 8 | 10 |
| X=FIB-BF, $N = 30$, $F = 10$ | (cf. Figure 13a) | 0.69 | 0.5 | 0.392 | 0.322 |
| X=FIB-BF, $N = 50$, $F = 15$ | (cf. Figure 13b) | 0.74 | 0.552 | 0.44 | 0.366 |
| X=FIB-BF, $N = 50$, $F = 10$ | | 0.686 | 0.497 | 0.389 | 0.32 |

In summary, FIB-BF saves more memory compared to FIB and FIB-Hash if the prefixes are distributed equally among the faces. However, FIB-Hash is more flexible because FIB-Hash also saves memory if prefixes are not distributed equally among the faces. Furthermore, FIB-Hash allows removal of prefixes in contrary to FIB-BF. FIB-Hash and FIB-BF/CBF always save memory compared to FIB if the prefixes are long. This is obvious because FIB-Hash and FIB-BF/CBF are independent of the prefix length. In Table 9 gives a comprehensive comparison of our proposed solutions.

Table 9 Comprehensive comparison of the proposed solutions.

| FIB Type | Mapping characteristic | Element removal | Deterministic (no false-positives) | Compression ratio (typical cases ) |
|---|---|---|---|---|
| FIB-Hash | no limitations | yes | yes | $\sim {}^{1}\!/_{3}$ |
| FIB-BF | n-1 | no | no | $\sim {}^{1}\!/_{4}$ |
| FIB-CBF | n-1 | yes | no | $\sim 0$ |

In Section 4 we suggested a 1-1-mapping for our IoT solution which means one prefix per face. When using a Bloom-Filter solution for the FIB this means that a Bloom-Filter contains only one element. Bloom-Filters reach their full potential if they store multiple elements and therefore, we suggest a hash based solution for the FIB in our IoT setup like FIB-Hash to save memory.

# 7. CONCLUSION AND OUTLOOK

In this work, we present memory efficient solutions for the forwarding information base for content-centric networking. We also identify requirements (for prefix to face mappings) under which our proposed solutions work best. Under the assumption, that our FIB follows an n-1-mapping, which means that we have many prefixes and few faces and prefixes are equally distributed among faces, the Bloom-Filter solution named FIB-BF works best.

Unfortunately, Bloom-Filters do not allow the removal of elements, which is a severe drawback in dynamic content-centric networking scenarios where FIB entries are modified at runtime. To circumvent this issue, counting Bloom-Filters are used (FIB-CBF). FIB-CBF only saves significant memory compared to our standard FIB solution with less than 11 faces or if prefixes become considerable long (more than 15 byte). FIB-CBF is always worse regarding the memory consumption than FIB-BF and our FIB solution with hashing, named FIB-Hash. The memory consumption of FIB-BF and FIB-CBF increases with increasing number of faces.

The size of FIB-Hash increases logarithmic with increasing number of prefixes and increasing number of faces and therefore, reduces memory consumption independent of the mapping between prefixes and faces. FIB-Hash compresses the FIB by about ${}^{1}\!/_{3}$ and FIB-BF by about ${}^{1}\!/_{4}$ in typical cases.

In future work, we will implement our proposed solutions on a wireless sensor node platform for the IoT. With real IoT applications we will provide a comparative evaluation of the computation time for our suggested approaches. We will also investigate how we can apply hashing and Bloom-Filter in reducing the size of other CCN data structures like the PIT and the CS. Another issue when working with wireless resource constrained IoT devices is the limited radio bandwidth. In future, we will investigate how hashing and Bloom-Filter approaches can be applied to save bandwidth without altering the concepts of CCN. For example, it is an open question if there are hashing or Bloom-Filter approaches which reduce the message size while preserving the original CCN name comparison.

## ACKNOWLEDGEMENTS

This work was funded by the Federal Ministry for Economic Affairs & Energy (035SX361C, BOSS). Horst Hellbrück is adjunct professor at the Institute of Telematics of University of Lübeck.

**Authors**

Torsten Teubler is a research associate and PhD student at Lübeck University of Applied Sciences in Lübeck, Germany. He received his diploma in computer science in 2009. He worked with several research projects ranging from Wireless Sensor Networks (WSN), Internet of Things (IoT), Content-Centric Networking (CCN), and Autonomous Underwater Vehicles (AUV) with focus on networking, protocols, and expert systems. His research interests are content-centric networking with focus on the internet of things and resource constrained devices.

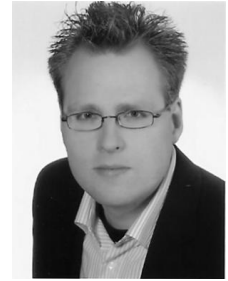

Dennis Pfisterer is a professor at DHBW Baden-Wuerttemberg Cooperative State University, Germany and adjunct professor at the University of Lübeck, Germany. His research interests include Internet of Things (IoT), Cyber-Physical Systems (CPS), Peer-to-Peer Networks (P2P), Semantic Web and IT security. He received his PhD (2007) and habilitation (2012) from the University of Lübeck, Germany where he worked on wireless sensor networks and the Internet of Things. He was research assistant at the European Media Laboratory (EML), Heidelberg, Germany and the Braunschweig Institute of Technology, Germany as well as soft- and hardware developer at the coalesenses GmbH, Lübeck, Germany. More information is available at www.dennis-pfisterer.de.

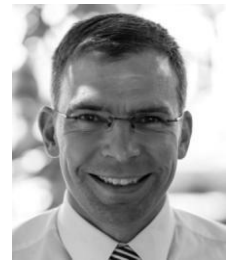

Horst Hellbrück is a professor at the University of Applied Sciences in Lübeck, Germany and adjunct professor at the University of Lübeck, Germany. His research interests are modern structures like wireless mobile networks and sensor networks. Of special interest are sensor networks in various application fields from medical applications to underwater technologies. He received his Diploma in Electrical Engineering in 1994 at the University of Saarland. He worked as a Software Engineer from 1994-1998 at Dräger in Lübeck and in International Product Marketing 1998-2000 at eupec in Warstein. He joined the International University in Germany, Bruchsal in 2000 and received his Doctorate in Computer Science 2004 at Technical University in Braunschweig in the field of ad-hoc networking. Before starting his professorship in communication systems and distributed systems at the university of applied sciences in 2008 he hold a position as post doc researcher at the Institute of Telematics at the University of Lübeck, Germany. Since December 2013 he is the head of Center of Excellence CoSA and has been appointed as adjunct professor since 2016 at the University of Lübeck at the Institute of Telematics.

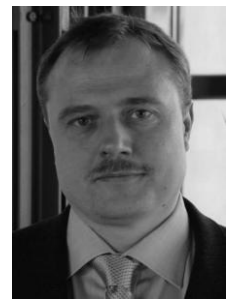